\documentclass[aps,prb,twocolumn,superscriptaddress,amsmath]{revtex4}

\usepackage{graphicx}
\usepackage{bm}

\begin{document}

\title{Temperature Dependence of
the Superfluid Density
in a Noncentrosymmetric Superconductor}

\author{N. Hayashi}
  \affiliation{
   Institut f\"ur Theoretische Physik,
   ETH-H\"onggerberg,
   CH-8093 Z\"urich, Switzerland
  }
\author{K. Wakabayashi}
  \affiliation{
   Institut f\"ur Theoretische Physik,
   ETH-H\"onggerberg,
   CH-8093 Z\"urich, Switzerland
  }
  \affiliation{
   Department of Quantum Matter Science,
   Graduate School of Advanced Sciences of Matter (ADSM),
   Hiroshima University, Higashi-Hiroshima 739-8530, Japan
  }
\author{P. A. Frigeri}
  \affiliation{
   Institut f\"ur Theoretische Physik,
   ETH-H\"onggerberg,
   CH-8093 Z\"urich, Switzerland
  }
\author{M. Sigrist}
  \affiliation{
   Institut f\"ur Theoretische Physik,
   ETH-H\"onggerberg,
   CH-8093 Z\"urich, Switzerland
  }

\date{\today}

\begin{abstract}
   For a noncentrosymmetric superconductor such as CePt$_3$Si,
we consider a Cooper pairing model with
a two-component order parameter
composed of spin-singlet and spin-triplet pairing components.
   We calculate
the superfluid density tensor
in the clean limit
on the basis of the quasiclassical theory of superconductivity.
   We demonstrate
that such a pairing model accounts for
an experimentally observed feature of the temperature dependence
of the London penetration depth in CePt$_3$Si,
i.e., line-node-gap behavior at low temperatures.
\end{abstract}


\maketitle

\section{Introduction}
   Much attention has been focused on
the superconductivity
in systems without inversion symmetry
({\it e.g.},
Refs.\ \onlinecite{edelstein,gorkov,paolo1,samokhin,sergienko,sergienko2},
and references therein).
   Recently,
CePt$_3$Si was found to be a heavy fermion
superconductor without inversion symmetry
in the crystal structure.\cite{bauer,bauer3,bauer2,bauer4}
 This motivates more detailed studies of the superconductivity in
noncentrosymmetric systems.
   The lack of an inversion center in
the crystal lattice induces
antisymmetric spin-orbit coupling\cite{rashba,dresselhaus}
responsible
for a mixing of spin-singlet and spin-triplet Cooper pairings.\cite{gorkov}
   In CePt$_3$Si,
this mixing of the pairing channels with different parity
may result in unusual properties
of experimentally observed quantities
such as
a very high upper critical field
$H_{\mathrm c2}$
which exceeds
the paramagnetic limit,\cite{bauer,bauer3,bauer2,bauer4,yasuda}
and the simultaneous appearance of 
a coherence peak feature
in the NMR relaxation rate $T_1^{-1}$
and low-temperature powerlaw behavior suggesting line-nodes
in the quasiparticle gap\cite{yogi,yogi2,yogi3,bauer3,bauer2,bauer4}
(see also Ref.\ \onlinecite{ueda}).
 The presence of line-nodes in the gap of CePt$_3$Si
is also indicated  by measurements
of
the thermal conductivity\cite{izawa}
and
the London penetration depth.\cite{bauer2,bonalde}

   In CePt$_3$Si,
the superconductivity coexists with
an antiferromagnetic
phase\cite{bauer,bauer3,bauer2,bauer4,yogi,yogi2,yogi3,metoki,amato}
(see also Ref.\ \onlinecite{ishikawa}).
   Generally one may have to include this aspect
when the low temperature thermodynamics is analyzed in this material.
The London penetration depth, however, 
which is entirely connected with the superfluid density,
contains exclusively
the information on the superconductivity,
and provides for this reason a very suitable probe of the low-energy
spectrum of the quasiparticles associated with the superconducting
gap topology.   Experimental measurements of the London penetration depth
on polycrystalline and powder samples
are reported in Refs.\ \onlinecite{bauer2,bonalde}.
   We note that
CePt$_3$Si is an extreme type-II superconductor
with the Ginzburg-Landau parameter $\kappa \simeq 140$,\cite{bauer,bauer2}
and the nonlocal effect can be safely neglected.

   For a noncentrosymmetric superconductor such as CePt$_3$Si,
we will consider a Cooper pairing model with
an order parameter consisting  of spin-singlet and spin-triplet pairing components.
Based on the same pairing model,
we previously investigated
the nuclear spin-lattice relaxation rate
to explain
peculiarities observed
in $T_1^{-1}$.\cite{hayashi05-1,hayashi05-2,samokhin2}
   In this paper, 
we calculate the superfluid density
and demonstrate
that this pairing model gives simultaneously an explanation of
the powerlaw temperature dependence
of the penetration depth at low temperatures
in CePt$_3$Si.

   This paper is organized as follows.
   In Sec.\ II,
we describe
the electronic structure of the system without inversion symmetry
and our pairing model.
   In Sec.\ III,
the equations for calculating the superfluid density
are formulated.
   The numerical results
are shown in Sec.\ IV.
   The summary is given in Sec.\ V.
   In the appendix,
we describe the derivation of the quasiclassical Green functions
used to compute the superfluid density for the present pairing model.

\section{System without Inversion Symmetry}
   We base our analysis on a system considered
in Ref.~\onlinecite{paolo1},
where
the lack of inversion symmetry
is incorporated through
the
antisymmetric
Rashba-type
spin-orbit coupling\cite{paolo1,paolo2,paolo2-2,paolo3,kaur}
\begin{equation}
\sum_{{\bm k},\eta,\eta'}
\alpha {\bm g}_k \cdot {\hat {\bm \sigma}}_{\eta\eta'}
c^{\dagger}_{{\bm k}\eta} c_{{\bm k}\eta'},
\label{eq:SO}
\end{equation}
with
\begin{equation}
{\bm g}_k
= \sqrt{\frac{3}{2}} \frac{1}{k_{\rm F}} \bigl(-k_y, k_x, 0 \bigr).
\label{eq:gk_SO}
\end{equation}
   Here,
${\hat {\bm \sigma}}=({\hat \sigma}_x,{\hat \sigma}_y,{\hat \sigma}_z)$
is the vector consisting of the Pauli matrices,
$c^{\dagger}_{{\bm k}\eta}$ ($c_{{\bm k}\eta}$)
is the creation (annihilation) operator
for the quasiparticle state with momentum ${\bm k}$ and spin $\eta$.
   We use units in which $\hbar = k_{\rm B} = 1$.
   $\alpha$ ($>0$) denotes the strength of the spin-orbit coupling.
   The antisymmetric vector ${\bm g}_k$
(${\bm g}_{-k} = -{\bm g}_k$)
is determined by symmetry arguments
and is normalized as
$\langle {\bm g}_k^2 \rangle_0 =1$.\cite{paolo1,paolo2}
   $k_{\rm F}$ is the Fermi wave number
and the brackets $\langle \cdots \rangle_0$
denote the average over the Fermi surface
in the case of $\alpha=0$.

   Generally we may classify the basic pairing states
for a superconductor of
given crystal symmetry,
distinguishing
the spin-singlet and spin-triplet states.\cite{paolo1,sergienko}
   A general argument by Anderson\cite{anderson}
shows that the inversion symmetry is a key
element for the realization of spin-triplet pairing states.
   Hence, the lack of inversion symmetry as in CePt$_3$Si
may be detrimental for spin-triplet pairing states.
   In other words, the presence of the antisymmetric spin-orbit coupling
would suppress spin-triplet pairing. 
   However, it has been shown by Frigeri {\it et al.}\cite{paolo1}
that 
the antisymmetric spin-orbit coupling
is not destructive to
the special spin-triplet state
with the ${\bm d}$ vector parallel to ${\bm g}_k$
(${\bm d}_k \parallel {\bm g}_k$).    
   Therefore, referring to
${\bm g}_k$ given in Eq.~(\ref{eq:gk_SO}),
we adopt
the $p$-wave pairing state with parallel ${\bm d}$ vector,
${\bm d}_k
=\Delta
(-{\tilde k}_y,{\tilde k}_x,0)$.\cite{d-vector}
Here,
the unit vector
${\tilde {\bm k}}
 =({\tilde k}_x,{\tilde k}_y,{\tilde k}_z)
=(\cos\phi \sin\theta, \sin\phi \sin\theta, \cos\theta)$.
   A further effect of the antisymmetric
spin-orbit coupling is the mixing of 
spin-singlet and spin-triplet
pairing components.\cite{gorkov}
   Interestingly, only the $s$-wave spin-singlet pairing state
(belonging to $A_{1g}$ representation of
crystal point group) mixes with 
the above $p$-wave spin-triplet pairing state
(for example, $d$-wave states cannot mix with
this $p$-wave state because of symmetry).\cite{paolo3,fujimoto}

   This parity-mixed pairing state is expressed by
the order parameter,
\begin{eqnarray}
{\hat \Delta}({\bm r}, {\tilde {\bm k}})
&=&                        \nonumber
\Bigl[
  \Psi({\bm r}) {\hat \sigma}_0
  + {\bm d}_k({\bm r}) \cdot {\hat {\bm \sigma}}
\Bigr]
i{\hat \sigma}_y                  \\
&=&                        \nonumber
\Bigl[
\Psi({\bm r}) {\hat \sigma}_0
 +
\Delta({\bm r})
\bigl(
  - {\tilde k}_y {\hat \sigma}_x
  + {\tilde k}_x {\hat \sigma}_y
\bigr)
\Bigr]
i{\hat \sigma}_y,   \\
\label{eq:OP-A2+s}
\end{eqnarray}
with the spin-singlet $s$-wave component $\Psi({\bm r})$ 
and
the ${\bm d}$ vector
${\bm d}_k({\bm r}) = \Delta({\bm r}) (-{\tilde k}_y,{\tilde k}_x,0)$.
   Here,
the vector ${\bm r}$ indicates the real-space coordinates,
and ${\hat \sigma}_0$ is the unit matrix in the spin space.
   While this spin-triplet part alone has point nodes,
the pairing state of Eq.\ (\ref{eq:OP-A2+s})
can possess line nodes in a gap
as a result of the combination with
the $s$-wave component.\cite{hayashi05-1,paolo3,sergienko3}
In this paper,
we choose the isotropic $s$-wave pairing as $\Psi$ for simplicity.

\section{Quasiclassical Formulation}
   We will calculate
the superfluid density
on the basis of
the quasiclassical theory of superconductivity.\cite{eilen,LO,serene}
   Following the spirit of the theory,\cite{serene}
in this study we assume
$|\Psi|, |\Delta|, \alpha \ll \varepsilon_{\mathrm F}$
($\varepsilon_{\mathrm F}$ is the Fermi energy).
   We consider the quasiclassical Green function ${\check g}$
which has
the matrix elements in Nambu (particle-hole) space
as
\begin{equation}
{\check g} ({\bm r}, {\tilde {\bm k}}, i\omega_n) =
-i\pi
\begin{pmatrix}
{\hat g} &
i{\hat f} \\
-i{\hat {\bar f} } &
-{\hat {\bar g} }
\end{pmatrix},
\label{eq:qcg}
\end{equation}
where
$\omega_n=\pi T (2n+1)$ is the Matsubara frequency
(with the temperature $T$ and the integer $n$).
   Throughout the paper,
{\it ``hat"} ${\hat \bullet}$ 
denotes the $2\times2$ matrix in the spin space,
and
{\it ``check"} ${\check \bullet}$ 
denotes the $4\times4$ matrix
composed of the $2\times2$ Nambu space
and the $2\times2$ spin space.

   The Eilenberger equation which includes
the spin-orbit coupling term
is given as\cite{hayashi05-1,schopohl80,rieck,choi,kusunose,eq-eilen}
\begin{equation}
i {\bm v}_{\mathrm F}(s) \cdot
{\bm \nabla}{\check g}
+ \bigl[ i\omega_n {\check \tau}_{3}
-\alpha {\check {\bm g}}_k \cdot {\check {\bm S}}
-{\check \Delta},
{\check g} \bigr]
=0,
\label{eq:eilen0}
\end{equation}
with
\begin{equation}
{\check \tau}_3 =
\begin{pmatrix}
{\hat \sigma}_0 &
0 \\
0 &
-{\hat \sigma}_0
\end{pmatrix},
\label{eq:tau3}
\end{equation}
\begin{equation}
{\check {\bm S}} =
\begin{pmatrix}
{\hat {\bm \sigma}} &
0 \\
0 &
{\hat {\bm \sigma}}^{tr}
\end{pmatrix},
\quad
{\hat {\bm \sigma}}^{tr}
=-{\hat \sigma}_y {\hat {\bm \sigma}} {\hat \sigma}_y,
\label{eq:spin}
\end{equation}
\begin{equation}
{\check {\bm g}}_k =
\begin{pmatrix}
{\bm g}_k {\hat \sigma}_0  &
0 \\
0 &
{\bm g}_{-k} {\hat \sigma}_0
\end{pmatrix}
=
\begin{pmatrix}
{\bm g}_k {\hat \sigma}_0 &
0 \\
0 &
-{\bm g}_k {\hat \sigma}_0
\end{pmatrix},
\end{equation}
\begin{equation}
{\bm g}_k
= \sqrt{\frac{3}{2}} \bigl(-{\tilde k}_y, {\tilde k}_x, 0 \bigr),
\label{eq:gk_SO-2}
\end{equation}
\begin{equation}
{\check  \Delta} =
\begin{pmatrix}
0 &
{\hat \Delta} \\
-{\hat \Delta}^\dagger &
0
\end{pmatrix}.
\label{eq:Delta}
\end{equation}
   Here,
${\bm v}_{\mathrm F}(s)$ is the Fermi velocity,
the variable $s$
indicates the position
on the Fermi surfaces,
and
the commutator
$[{\check a},{\check b}]={\check a}{\check b}-{\check b}{\check a}$.
   The Eilenberger equation is supplemented
by the normalization condition\cite{eilen,schopohl80}
\begin{equation}
   {\check g}^2 =
   -\pi^2 {\check 1},
\label{eq:normalization}
\end{equation}
where ${\check 1}$ is the $4\times4$ unit matrix.
   Because CePt$_3$Si is a clean superconductor,\cite{bauer,bauer2}
we neglect the impurity effect.

   To obtain an expression for the superfluid density,
we follow the procedure developed by
Choi and Muzikar.\cite{choi87,choi88,choi89,muzikar95,choi90}
   We consider a system in which
a uniform supercurrent flows
with the velocity ${\bm v}_s$,
and the gap function (\ref{eq:OP-A2+s})
has the ${\bm r}$ dependence as\cite{choi87,choi88,choi89,muzikar95,choi90}
\begin{eqnarray}
{\hat \Delta}({\bm r},{\tilde {\bm k}})
&=&                        \nonumber
{\hat \Delta}'({\tilde {\bm k}})
e^{i2M {\bm v}_s \cdot {\bm r} }                  \\
&=&                        \nonumber
\Bigl[
\Psi {\hat \sigma}_0
 +
\Delta
\bigl(
  - {\tilde k}_y {\hat \sigma}_x
  + {\tilde k}_x {\hat \sigma}_y
\bigr)
\Bigr]
i{\hat \sigma}_y
e^{i2M {\bm v}_s \cdot {\bm r} },   \\
\end{eqnarray}
and accordingly
\begin{eqnarray}
\Psi({\bm r})
=
\Psi
e^{i2M {\bm v}_s \cdot {\bm r} },
\quad
\Delta({\bm r})
=
\Delta
e^{i2M {\bm v}_s \cdot {\bm r} }.
\end{eqnarray}
   The bare electron mass is denoted by $M$.
   The matrix elements of the Green function (\ref{eq:qcg})
are expressed as
\begin{subequations}
\begin{eqnarray}
{\hat g}({\bm r},{\tilde {\bm k}},i\omega_n)
=
{\hat g}'({\tilde {\bm k}},i\omega_n),
\label{eq:g-prime}
\end{eqnarray}
\begin{eqnarray}
{\hat f}({\bm r},{\tilde {\bm k}},i\omega_n)
=
{\hat f}'({\tilde {\bm k}},i\omega_n)
e^{i2M {\bm v}_s \cdot {\bm r} },
\end{eqnarray}
\begin{eqnarray}
{\hat {\bar f} }({\bm r},{\tilde {\bm k}},i\omega_n)
=
{\hat {\bar f}' }({\tilde {\bm k}},i\omega_n)
e^{-i2M {\bm v}_s \cdot {\bm r} },
\end{eqnarray}
\begin{eqnarray}
{\hat {\bar g} }({\bm r},{\tilde {\bm k}},i\omega_n)
=
{\hat {\bar g}' }({\tilde {\bm k}},i\omega_n).
\end{eqnarray}
\end{subequations}
   The Eilenberger equation (\ref{eq:eilen0})
is rewritten
in a form without the ${\bm r}$ dependence as
\begin{equation}
\bigl[ i(\omega_n +q) {\check \tau}_{3}
-\alpha {\check {\bm g}}_k \cdot {\check {\bm S}}
-{\check \Delta}',
{\check g}' \bigr]
=0,
\label{eq:eilen-SFD}
\end{equation}
with
\begin{equation}
q=iM {\bm v}_{\mathrm F}(s) \cdot {\bm v}_s,
\end{equation}
\begin{equation}
{\check g}'=
-i\pi
\begin{pmatrix}
{\hat g}' &
i{\hat f}' \\
-i{\hat {\bar f}' } &
-{\hat {\bar g}' }
\end{pmatrix},
\end{equation}
\begin{equation}
{\check  \Delta}' =
\begin{pmatrix}
0 &
{\hat \Delta}' \\
-{\hat \Delta}^{'\dagger} &
0
\end{pmatrix}.
\label{eq:Delta2}
\end{equation}

   We obtain the following Green functions
from the Eilenberger equation
and the normalization condition (see Appendix),
\begin{subequations}
\begin{eqnarray}
{\hat g}'
&=&
g_{\rm I} {\hat \sigma}_{\rm I}
+
g_{\rm II} {\hat \sigma}_{\rm II},
\label{eq:g-prime2}
  \\
{\hat f}'
&=&
\bigl(
f_{\rm I} {\hat \sigma}_{\rm I}
+
f_{\rm II} {\hat \sigma}_{\rm II}
\bigr) i{\hat \sigma}_y,
\end{eqnarray}
\label{eq:Green-uniform}
\end{subequations}
%
%
with the matrices ${\hat \sigma}_{\rm I}$ and ${\hat \sigma}_{\rm II}$
defined by\cite{edelstein,paolo3,mineev}
\vspace{-1mm}
\begin{eqnarray}
{\hat \sigma}_{\rm I,II} = \frac{1}{2}
\Bigl(
{\hat \sigma}_0
   \pm {\bar {\bm g}}_k \cdot {\hat {\bm \sigma}}
\Bigr),
\quad
{\bar {\bm g}}_k
= (-{\bar k}_y,{\bar k}_x,0).
\vspace{-1mm}
\end{eqnarray}
%
%
%
%
%
   Here,
${\bar {\bm k}} = ({\bar k}_x,{\bar k}_y,0) = (\cos\phi,\sin\phi,0)$,
and
\begin{subequations}
\label{eq:g-elements}
\begin{eqnarray}
g_{\rm I}
=
    \frac{ \omega_n +q }{B_{\rm I}},
\quad
g_{\rm II}
=
    \frac{ \omega_n +q }{B_{\rm II}},
\label{eq:g}
\end{eqnarray}
\begin{eqnarray}
f_{\rm I}
=
    \frac{ \Psi + \Delta\sin\theta }{B_{\rm I}},
\label{eq:f_0}
\quad
f_{\rm II}
=
    \frac{ \Psi - \Delta\sin\theta }{B_{\rm II}}.
\label{eq:f}
\end{eqnarray}
\end{subequations}
   The denominators
$B_{\rm I}$ and $B_{\rm II}$ are given as
\begin{subequations}
\begin{eqnarray}
B_{\rm I}
=
\pm \sqrt{
      (\omega_n +q)^2 
      +|\Psi + \Delta\sin\theta|^2
    },
\end{eqnarray}
\begin{eqnarray}
B_{\rm II}
=
\pm \sqrt{
      (\omega_n +q)^2 
      +|\Psi - \Delta\sin\theta|^2
    },
\end{eqnarray}
\end{subequations}
   and the signs in front of the square root
are determined by the conditions
\begin{subequations}
\begin{eqnarray}
{\rm sgn}
\bigl(
  {\rm Re} \{ g_{\rm I} \}
\bigr)
=
{\rm sgn}
\bigl(
  {\rm Re} \{ \omega_n \}
\bigr),   \\
{\rm sgn}
\bigl(
  {\rm Re} \{ g_{\rm II} \}
\bigr)
=
{\rm sgn}
\bigl(
  {\rm Re} \{ \omega_n \}
\bigr).
\end{eqnarray}
\end{subequations}

   The Green functions labeled by the indices I and II belong to
the two distinct Fermi surfaces which are split
by the lifting of the spin-degeneracy
due to the spin-orbit coupling.
   The density of states
on those two Fermi surfaces is different
from each other in general.
   We define
the density of states (the Fermi velocity)
as
$N_{\rm I,II}$ (${\bm v}_{\rm I,II}$)
on the Fermi surfaces I and II.
   We also define
a parameter $\delta$ ($-1<\delta<1$) which parameterizes
the difference in the density of states,
\begin{eqnarray}
\delta =
\frac{ N_{\rm I}-N_{\rm II} }{2 N_0},
\label{eq:delta}
\end{eqnarray}
where
$2 N_0= N_{\rm I}+N_{\rm II}$.

   The supercurrent ${\bm J}$ is
composed of the regular part, $-i\pi {\hat g}$, of the Green function
in Eq.\ (\ref{eq:qcg}).\cite{serene}
   ${\bm J}$ is expressed by
\begin{eqnarray}
{\bm J}
&=&
T \sum_{\omega_n} \int ds N_{\rm F}(s) {\bm v}_{\rm F}(s)
{\rm tr}
\bigl[
   {\hat \sigma}_0 (-i\pi {\hat g})
\bigr]
\nonumber  \\
&=&
\frac{\pi T}{i}
\sum_{\omega_n}
\Bigl[
   N_{\rm I}
   \bigl\langle
      {\bm v}_{\rm I}
      g_{\rm I}
   \bigr\rangle
   +
   N_{\rm II}
   \bigl\langle
      {\bm v}_{\rm II}
      g_{\rm II}
   \bigr\rangle
\Bigr],
\label{eq:J}
\end{eqnarray}
where
$N_{\rm F}(s)$ is the density of states
at the position $s$
on the Fermi surfaces,
``tr" means the trace in the spin space,
and
the brackets $\langle \cdots \rangle$
denote the average over each Fermi surface.
   In Eq.\ (\ref{eq:J}), we have referred to
Eqs.\ (\ref{eq:g-prime}) and (\ref{eq:g-prime2}).
   In order to calculate
the superfluid density tensor $\rho_{ij}$
($J_i = \rho_{ij} v_{sj}$),
we expand $g_{\rm I,II}$ in Eq.\ (\ref{eq:g})
up to first order in ${\bm v}_s$ (or $q$),
and substitute them into Eq.\ (\ref{eq:J}).
   The expression for $\rho_{ij}$
is then obtained as\cite{gk}
\begin{eqnarray}
\rho_{ij}
&=&        \nonumber
M\pi T
\sum_{\omega_n}
\Biggl[
   N_{\rm I}
   \Biggl\langle
     \frac{
       v_{{\rm I} i}
       v_{{\rm I} j}
       |\Psi + \Delta\sin\theta|^2
     }
     {
       \bigl(
         \omega_n^2 
         +|\Psi + \Delta\sin\theta|^2
       \bigr)^{3/2}
     }
   \Biggr\rangle
\\ & & {}
   +
   N_{\rm II}
   \Biggl\langle
     \frac{
      v_{{\rm II} i}
      v_{{\rm II} j}
       |\Psi - \Delta\sin\theta|^2
     }
     {
       \bigl(
         \omega_n^2 
         +|\Psi - \Delta\sin\theta|^2
       \bigr)^{3/2}
     }
   \Biggr\rangle
\Biggr].
\label{eq:rho}
\end{eqnarray}

   Now, to compute the superfluid density tensor $\rho_{ij}$
in Eq.\ (\ref{eq:rho}),
we need to assume a model of the Fermi surfaces.
  For the shape of the Fermi surfaces,
we adopt the spherical Fermi surface for simplicity,
and thus
${\bm v}_{\rm I,II}=v_{\rm I,II} {\tilde {\bm k}}
=v_{\rm I,II}(\cos\phi \sin\theta, \sin\phi \sin\theta, \cos\theta)$
and
the Fermi-surface average
$\langle \cdots \rangle
=(1/4\pi)\int_0^{2\pi} d\phi
 \int_0^\pi d\theta \sin\theta \cdots$.\cite{FS-SOC}
   This spherical approximation is justified for CePt$_3$Si,
since measurements of $H_{\mathrm c2}$ give
evidence for a nearly isotropic mass tensor.\cite{bauer,yasuda}
   We furthermore set the two Fermi velocities equal,\cite{yip}
\begin{eqnarray}
v_{\rm I} = v_{\rm II} \equiv v.
\label{eq:model-vF}
\end{eqnarray}
   From Eq.\ (\ref{eq:rho}),
the components of
the superfluid density tensor are obtained as
\begin{eqnarray}
\rho_{xx}
&=&        \nonumber
(M v^2 N_0)
2\pi T
\sum_{\omega_n >0}
\\ & & {}        \nonumber
\times
\frac{1}{2}
\Biggl[
   C_{\rm I}(\delta)
   \int_0^\pi \frac{d\theta}{2}
     \frac{
       |\Psi + \Delta\sin\theta|^2
       \sin^3 \theta
     }
     {
       \bigl(
         \omega_n^2 
         +|\Psi + \Delta\sin\theta|^2
       \bigr)^{3/2}
     }
\\ & & {}        \nonumber
   +
   C_{\rm II}(\delta)
   \int_0^\pi \frac{d\theta}{2}
     \frac{
       |\Psi - \Delta\sin\theta|^2
       \sin^3 \theta
     }
     {
       \bigl(
         \omega_n^2 
         +|\Psi - \Delta\sin\theta|^2
       \bigr)^{3/2}
     }
\Biggr],   \\
\label{eq:rho_xx}
\end{eqnarray}
\begin{eqnarray}
\rho_{zz}
&=&        \nonumber
(M v^2 N_0)
2\pi T
\sum_{\omega_n >0}
\\ & & {}        \nonumber
\times
\Biggl[
   C_{\rm I}(\delta)
   \int_0^\pi \frac{d\theta}{2}
     \frac{
       |\Psi + \Delta\sin\theta|^2
       \sin\theta
       \cos^2 \theta
     }
     {
       \bigl(
         \omega_n^2 
         +|\Psi + \Delta\sin\theta|^2
       \bigr)^{3/2}
     }
\\ & & {}        \nonumber
   +
   C_{\rm II}(\delta)
   \int_0^\pi \frac{d\theta}{2}
     \frac{
       |\Psi - \Delta\sin\theta|^2
       \sin\theta
       \cos^2 \theta
     }
     {
       \bigl(
         \omega_n^2 
         +|\Psi - \Delta\sin\theta|^2
       \bigr)^{3/2}
     }
\Biggr],   \\
\label{eq:rho_zz}
\end{eqnarray}
$\rho_{xy}=\rho_{yx}=0$,
$\rho_{zx}=\rho_{xz}=\rho_{zy}=\rho_{yz}=0$,
and
$\rho_{yy}=\rho_{xx}$.
   Here, the weighting factors
$C_{\rm I,II}$ in the case of the model of Eq.\ (\ref{eq:model-vF})
are given as
\begin{eqnarray}
C_{\rm I}(\delta)=1 + \delta,
\quad
C_{\rm II}(\delta)=1 - \delta.
\label{eq:C_I-II}
\end{eqnarray}
   At zero temperature,
\begin{eqnarray}
\rho
&\equiv&        \nonumber
\rho_{xx}(T=0)=\rho_{zz}(T=0)
\\
&=&
\frac{2}{3}
M v^2 N_0,
\end{eqnarray}
where we have utilized a formula
for an arbitrary function~$F$,
$\lim_{T \rightarrow 0} 2\pi T \sum_{\omega_n} F(\omega_n)
=\int d\omega F(\omega)$.

   The gap equations
for the order parameters $\Psi$ and $\Delta$
are given by\cite{paolo3,gk}
\begin{eqnarray}
\Psi
&=&      \nonumber
\pi T
\sum_{|\omega_n| < \omega_{\mathrm c}}
\Bigl[
   \lambda_s 
   \bigl\langle
      f_+
   \bigr\rangle
   +
   \delta\lambda_s 
   \bigl\langle
      f_-
   \bigr\rangle
\\ & & {}
   +
   \lambda_m
   \bigl\langle
      \sin\theta
       f_-
   \bigr\rangle
   +
   \delta\lambda_m
   \bigl\langle
      \sin\theta
       f_+
   \bigr\rangle
\Bigr],
\label{eq:gap-singlet}
\end{eqnarray}
\begin{eqnarray}
\Delta
&=&      \nonumber
\pi T
\sum_{|\omega_n| < \omega_{\mathrm c}}
\Bigl[
   \lambda_t 
   \bigl\langle
      \sin\theta
      f_-
   \bigr\rangle
   +
   \delta\lambda_t 
   \bigl\langle
      \sin\theta
      f_+
   \bigr\rangle
\\ & & {}
   +
   \lambda_m
   \bigl\langle
       f_+
   \bigr\rangle
   +
   \delta\lambda_m
   \bigl\langle
       f_-
   \bigr\rangle
\Bigr],
\label{eq:gap-triplet}
\end{eqnarray}
where
\begin{eqnarray}
f_\pm=\frac{f_{\rm I} \pm f_{\rm II}}{2},
\end{eqnarray}
and $\omega_{\rm c}$ is the cutoff energy.
   The coupling constants
$\lambda_s$ and $\lambda_t$
result from the pairing interaction within each spin channel
($s$: singlet, $t$: triplet).
   $\lambda_m$ appears as a scattering of Cooper pairs between
the two channels, which is allowed
in a system without inversion symmetry.\cite{paolo3}

   In the limit $T \rightarrow T_{\mathrm c}$
($T_{\mathrm c}$ is the superconducting critical temperature),
the linearized gap equations allow us to determine
$\lambda_s$ and $\lambda_t$ by
\begin{eqnarray}
\lambda_s
=
\Bigl[
   \frac{1}{w}
   - \lambda_m
     \Bigl(
       \frac{2}{3\nu} + \frac{\pi\delta}{4}
     \Bigr)
\Bigr]
\Big/
\Bigl(
   1+\frac{\pi\delta}{4\nu}
\Bigr),
\end{eqnarray}
\begin{eqnarray}
\lambda_t
=
\Bigl[
   \frac{1}{w}
   - \lambda_m
     \Bigl(
       \nu + \frac{\pi\delta}{4}
     \Bigr)
\Bigr]
\Big/
\Bigl(
   \frac{2}{3}+\frac{\pi\delta\nu}{4}
\Bigr),
\end{eqnarray}
\begin{eqnarray}
w
=
\ln\Bigl(\frac{T}{ T_{\mathrm c} } \Bigr)
+ \sum_{0 \le n < (\omega_{\mathrm c}/\pi T -1)/2}   \frac{2}{2n+1},
\end{eqnarray}
\begin{eqnarray}
\nu
=
\frac{\Psi}{\Delta} \bigg|_{T \rightarrow T_{\mathrm c}-0^{+}},
\label{eq:nu}
\end{eqnarray}
when the parameters $\lambda_m$ and $\nu$
are given.

\section{Numerical Results}
   In this section, we will show the numerically evaluated results
for the superfluid densities $\rho_{xx}$ and $\rho_{zz}$.
   To calculate them
in Eqs.\ (\ref{eq:rho_xx}) and (\ref{eq:rho_zz}),
we need the temperature dependence of the order parameters
$\Psi$ and $\Delta$.
   We will use the order parameters obtained
from the gap equations
$\bigl[$Eqs.\ (\ref{eq:gap-singlet}) and (\ref{eq:gap-triplet})$\bigr]$
for the parameters $\omega_{\mathrm c}=100 T_{\mathrm c}$
and $\lambda_m=0.2$.
   This is a representative set of parameters.
   Different choices would not lead to 
qualitatively different results as long as the gap topology is not altered.
   We have calculated the superfluid densities also for $\lambda_m=0.1$
and obtained qualitatively same results.

\begin{figure}
\includegraphics[scale=0.19]{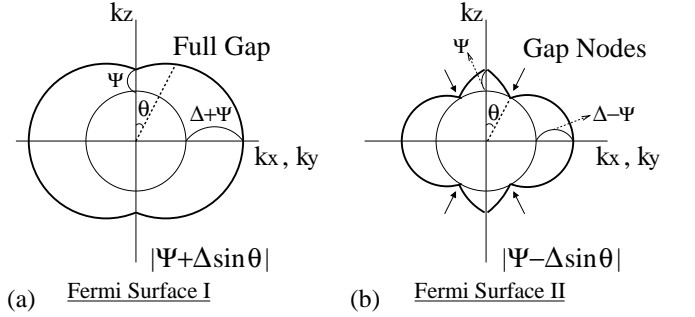}
\caption{
   Schematic figures of the gap structure
on the Fermi surfaces.
   (a) On the Fermi surface I,
the gap is $|\Psi+\Delta\sin\theta|$.
   (b) On the Fermi surface II,
the gap is $|\Psi-\Delta\sin\theta|$.
   In these figures, it is assumed that
both $\Psi$ and $\Delta$
are real and positive, and $\Psi<\Delta$.
}
\label{fig:1}
\end{figure}

   The Green functions in Eq.\ (\ref{eq:A-uniform-G})
are substituted into the gap equations.
   When solving the gap equations,
$\Delta$ is fixed to be real
without loss of generality, resulting in a real $\Psi$ as well.\cite{paolo3}
   Referring to Eq.\ (\ref{eq:f}),
we notice that
the superconducting gaps are $|\Psi+\Delta\sin\theta|$
and $|\Psi-\Delta\sin\theta|$
on the Fermi surfaces I and II, respectively.
   Such a gap structure can lead to
line nodes
on either Fermi surface I or II
(as shown in Fig.\ \ref{fig:1}).\cite{paolo3,hayashi05-1,sergienko3}
   When the signs of $\Psi$ and $\Delta$
are reverse to (same as) each other,
gap nodes appear on the Fermi surface I (II).
   The relative sign is controlled by
the parameters $\lambda_m$, $\delta$, and $\nu$
in the present formulation.
   In this paper,
we choose such parameters as $\Psi>0$ and $\Delta>0$,
so that gap nodes
($|\Psi-\Delta\sin\theta|=0$)
appear on the Fermi surface II (see Fig.\ \ref{fig:1}).
   Under this circumstance,
we can obtain stable order parameters $\Psi$ and $\Delta$ when
the difference in the density of states $\delta$
defined in Eq.\ (\ref{eq:delta})
is set as $\delta \agt -0.2$ for $\lambda_m=0.1$ and $0.2$.
   For $\delta \ge -0.2$,
stable $\Psi$ and $\Delta$ are obtained
when 
the singlet-to-triplet components ratio $\nu$
defined in Eq.\ (\ref{eq:nu})
is set as $\nu \agt 0.3$ ($\lambda_m=0.1$)
and $\nu \agt 0.5$ ($\lambda_m=0.2$).

\begin{figure}
\includegraphics[scale=0.6]{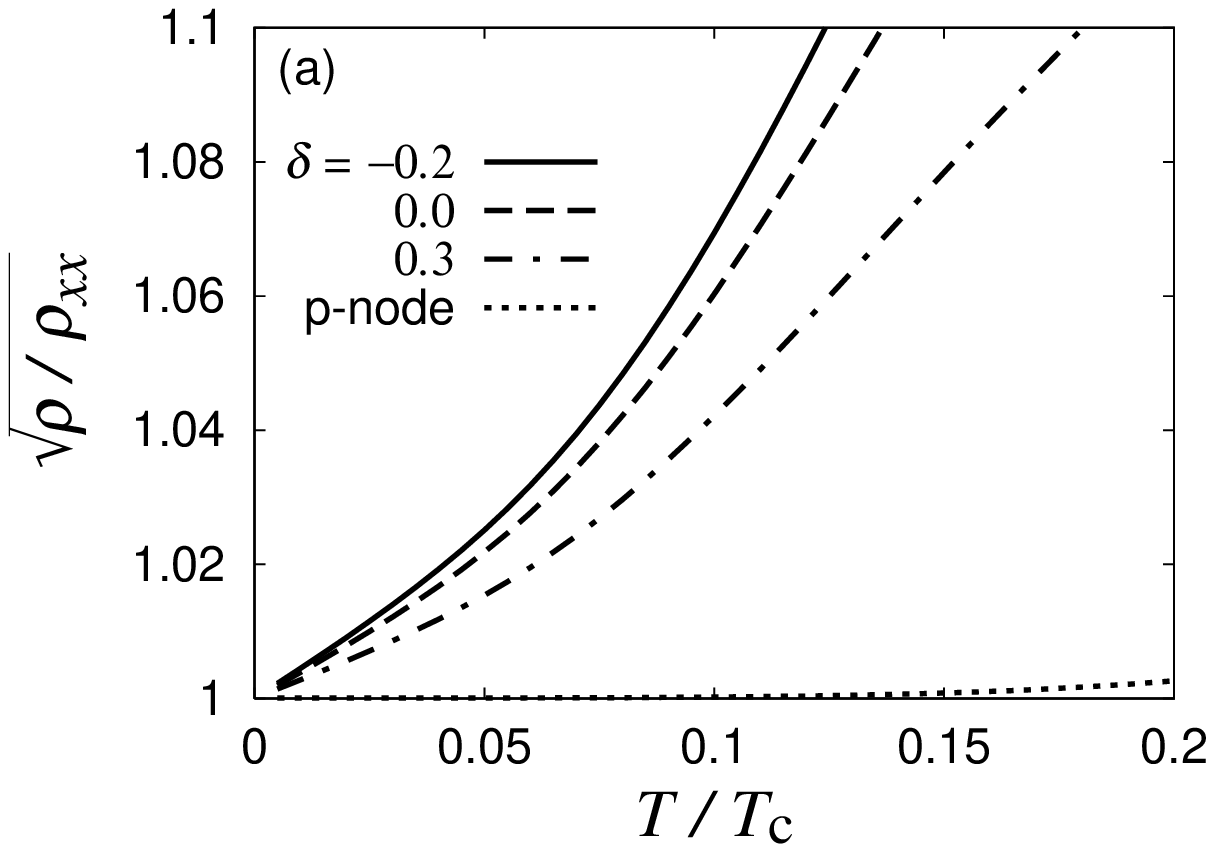}
\includegraphics[scale=0.6]{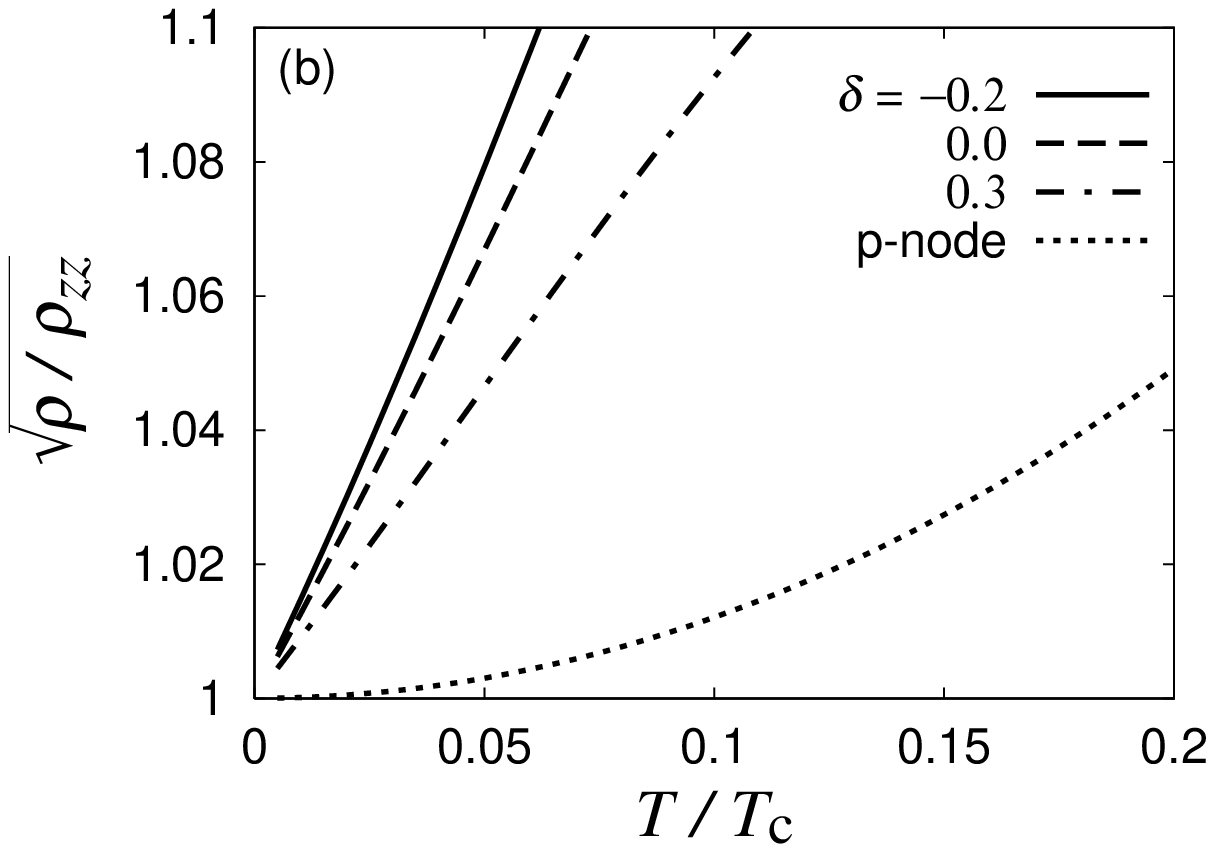}
\caption{
   Plots of
the reciprocal square root
of the superfluid densities vs.\ temperature,
$1/\sqrt{\rho_{xx}}$ (a) and $1/\sqrt{\rho_{zz}}$ (b),
for several values of the difference in the density of states $\delta$.
   The singlet-to-triplet components ratio $\nu$ is set as
$\nu=0.6$.
   For comparison,
dotted lines indicate
those for a point-node gap,
i.e., pure $p$-wave gap $|\Delta\sin\theta|$ ($\Psi=0$).
}
\label{fig:2}
\end{figure}

   In Fig.\ \ref{fig:2},
we show in a low-temperature region
the reciprocal square root
of the superfluid densities
$1/\sqrt{\rho_{xx}(T)}$ and $1/\sqrt{\rho_{zz}(T)}$,
which correspond to the London penetration depth $\lambda_{\mathrm L}(T)$.
   We set here the parameter $\nu=0.6$,
for which the gap nodes are line nodes
on the Fermi surface II.\cite{hayashi05-1}
   Indeed,
the data
exhibit the $T$-linear behavior at low temperatures,
indicating the existence of line nodes.
   For comparison, we also plot in Fig.\ \ref{fig:2}
the same quantities
calculated for a point-node gap (dotted line),
which is contrasting well with the line-node-gap behavior.
   The present results explain
the experimentally observed $T$-linear behavior of
$\lambda_{\mathrm L}(T)$ in CePt$_3$Si.\cite{bauer2,bonalde}
   We note here that
CePt$_3$Si is an extreme type-II superconductor\cite{bauer,bauer2}
and nonlocal effects can be neglected.

   Concerning the dependence on
$\delta$ the difference in the density of states
$\bigl[$Eq.\ (\ref{eq:delta})$\bigr]$,
we notice in Fig.\ \ref{fig:2} that
the smaller $\delta$ ($-1<\delta<1$),
the stronger temperature-dependence appears.
   Note the nearly temperature-independent 
behavior of the state with point-nodes (dotted line) at low temperatures
corresponding to a $ T^3 $ behavior.
This indicates weaker contributions of low-energy quasiparticles. 
 For the identical Fermi velocities on the two Fermi surfaces, 
the smaller $\delta$ ($-1<\delta<1$) leads to
a smaller weighting factor $C_{\rm I} =1+\delta$
and a larger $C_{\rm II}=1-\delta$.
   Therefore,
with decreasing $\delta$,
to the superfluid densities
the contribution of the fully-gapped Fermi surface I shrinks
with the decreasing weighting factor $C_{\rm I} =1+\delta$.
On the other hand, 
the contribution of the Fermi surface II with the gap nodes increases
because of the growing weighting factor $C_{\rm II} =1-\delta$ and 
 the effect of the line-node-gapped Fermi surface II
is enhanced by decreasing $\delta$ (Fig.\ \ref{fig:2}).

\begin{figure}
\includegraphics[scale=0.6]{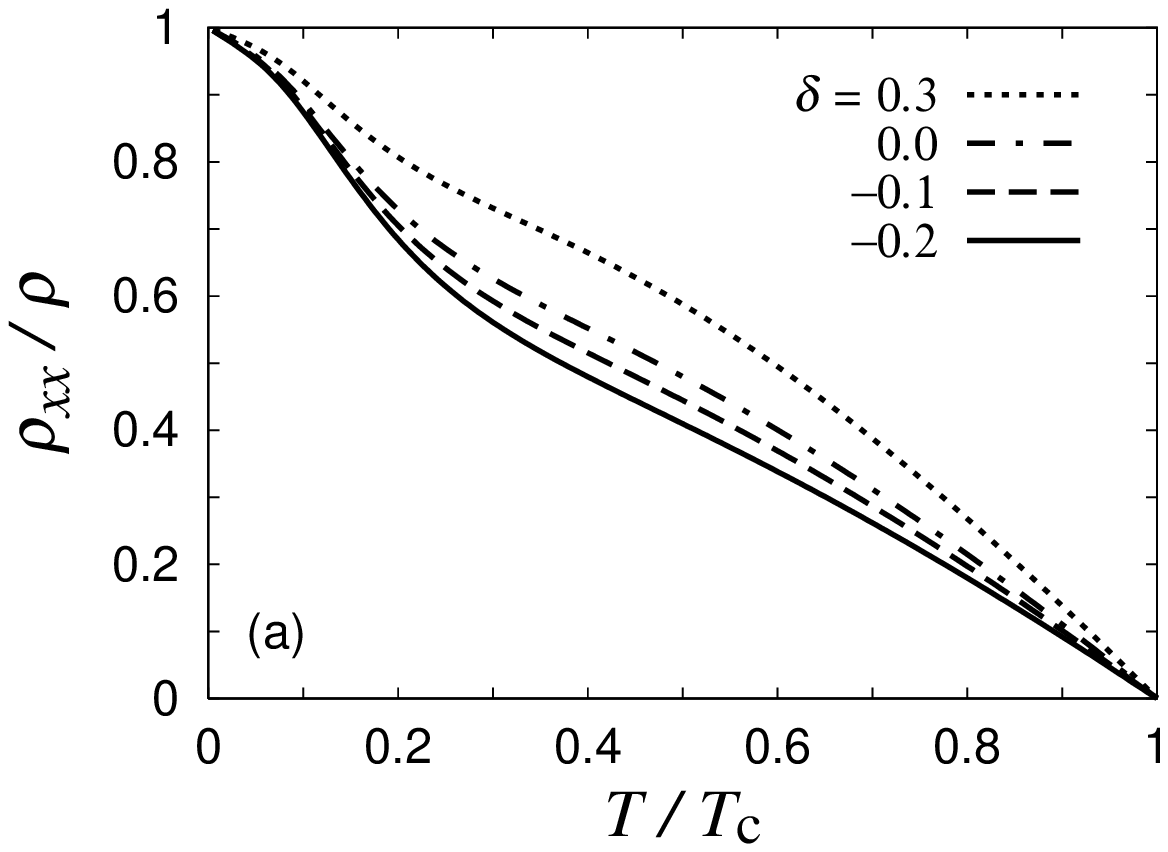}
\includegraphics[scale=0.6]{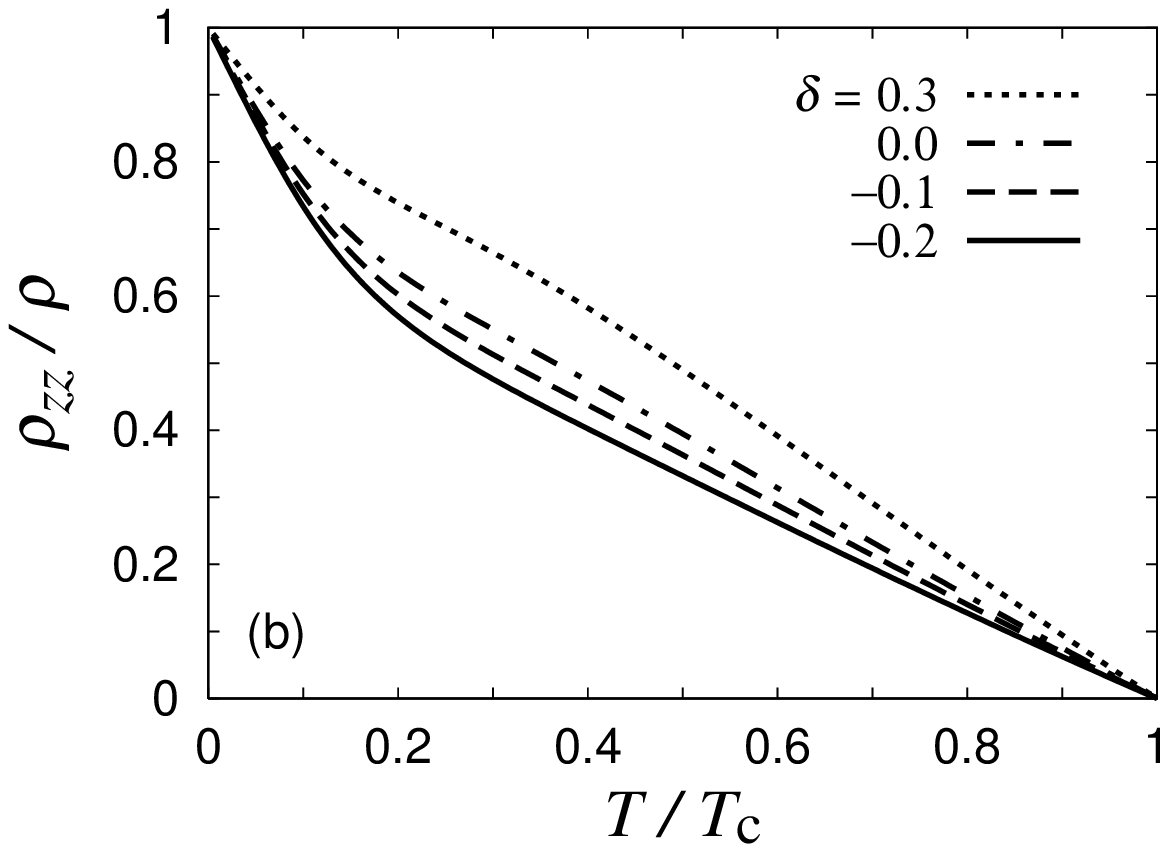}
\caption{
   The temperature dependence of
the superfluid densities,
$\rho_{xx}$ (a) and $\rho_{zz}$ (b),
for several values of the difference in the density of states $\delta$.
   The singlet-to-triplet components ratio $\nu$ is set as
$\nu=0.6$.
}
\label{fig:3}
\end{figure}

\begin{figure}
\includegraphics[scale=0.6]{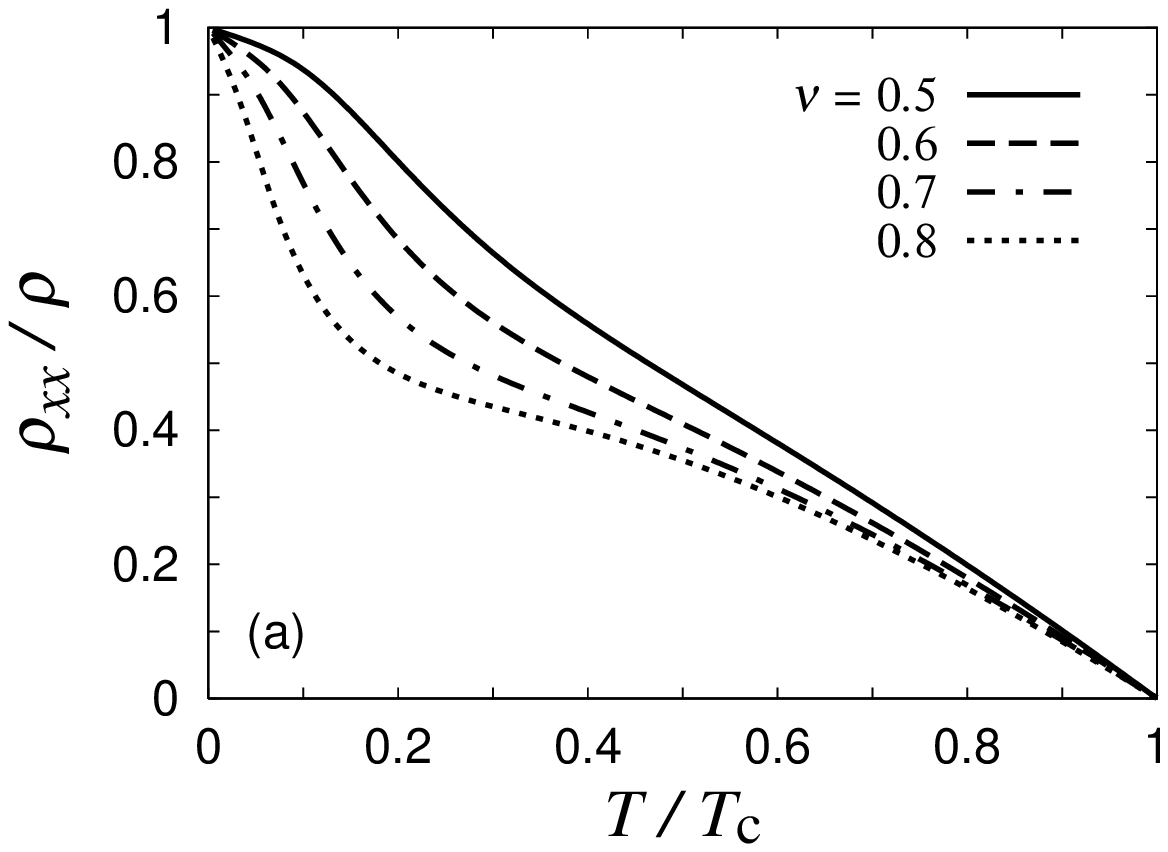}
\includegraphics[scale=0.6]{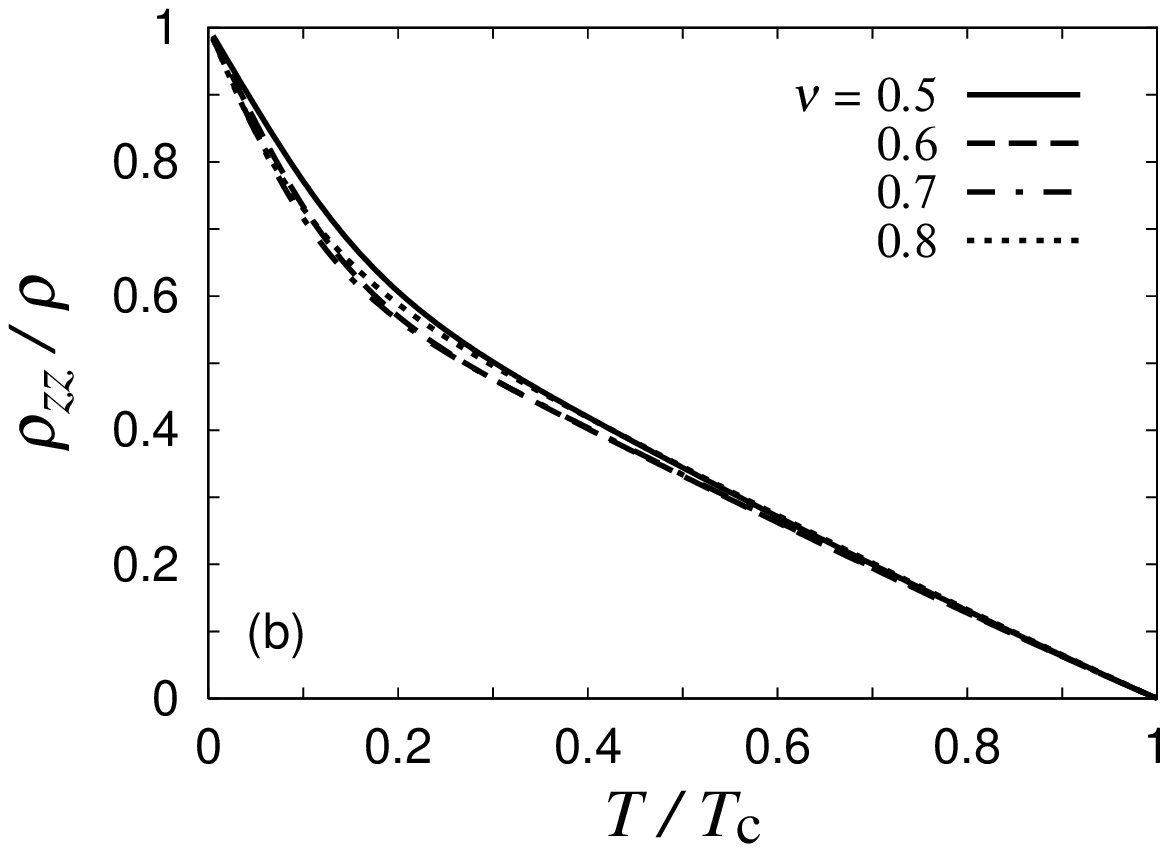}
\caption{
   The temperature dependence of
the superfluid densities,
$\rho_{xx}$ (a) and $\rho_{zz}$ (b),
for several values of the singlet-to-triplet components ratio $\nu$.
   The difference in the density of states is set as
$\delta=-0.2$.
}
\label{fig:4}
\end{figure}

   In Figs.\ \ref{fig:3} and \ref{fig:4},
we show
the superfluid densities
$\rho_{xx}(T)$ and $\rho_{zz}(T)$
to see the overall temperature dependence.
   The quantities correspond to $1/\lambda_{\mathrm L}^2(T)$.
   With decreasing $\delta$ ($-1<\delta<1$) in Fig.\ \ref{fig:3},
the curves deviate from an upper convex curve
and become gradually upper concave curves, namely
they deviate gradually from fully-gapped $s$-wave behavior
(i.e., $\rho_{ii}(T=0)-\rho_{ii}(T) \sim T^4$ in the $s$-wave case)
because of the same reason mentioned above
for the $\delta$ dependence in Fig.\ \ref{fig:2}.

   We show in Fig.\ \ref{fig:4}
the dependence on the singlet-to-triplet components ratio $\nu$
$\bigl[$Eq.\ (\ref{eq:nu})$\bigr]$.
   With increasing $\nu$ in Fig.\ \ref{fig:4}(a),
the curvature of the upper concave curves of $\rho_{xx}$
becomes larger.
   On the other hand,
the $\nu$ dependence of
$\rho_{zz}$ is weaker than that of $\rho_{xx}$,
as seen in Fig.\ \ref{fig:4}(b).
   The quantity $\rho_{xx}$
$\bigl[$Eq.\ (\ref{eq:rho_xx})$\bigr]$
senses the gap topology emphatically near the equator of the Fermi surfaces,
while
$\rho_{zz}$
$\bigl[$Eq.\ (\ref{eq:rho_zz})$\bigr]$
senses it near the poles.
   For the singlet-to-triplet components ratio $\nu \agt 0.5$,
the places (or the angles $\theta$) of gap nodes
at which $|\Psi-\Delta\sin\theta|=0$ (see Fig.\ \ref{fig:1})
are sufficiently away from the poles,
and approach gradually to
the equator on the Fermi surface II with increasing $\nu$.
   Therefore,
$\rho_{xx}$ ($\rho_{zz}$) is sensitive
(not sensitive)
to the change of $\nu$ for $\nu \ge 0.5$
as seen in Fig.\ \ref{fig:4}.

\begin{figure}
\includegraphics[scale=0.6]{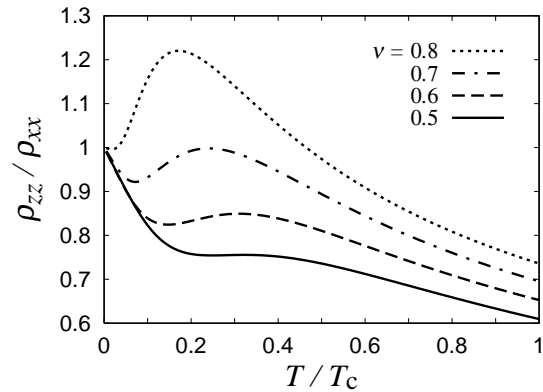}
\caption{
   Plots of the ratio
$\rho_{zz}/\rho_{xx}$
vs.\ temperature
for several values of the singlet-to-triplet components ratio $\nu$.
   The difference in the density of states is set as
$\delta=-0.2$.
}
\label{fig:5}
\end{figure}

   It is interesting to note the difference in the temperature dependence
between $\rho_{xx}$ and $\rho_{zz}$.
   In Fig.\ \ref{fig:5},
we plot the ratios
$\rho_{zz}/\rho_{xx}$
as functions of the temperature
for several values of $\nu$.
   They exhibit a nonmonotonic temperature dependence
in contrast with a monotonic one in the cases of
the axial state (point-node gap at the poles)
and the polar state (line-node gap at the equator)
shown in Ref.\ \onlinecite{choi89}.

\begin{figure}
\includegraphics[scale=0.6]{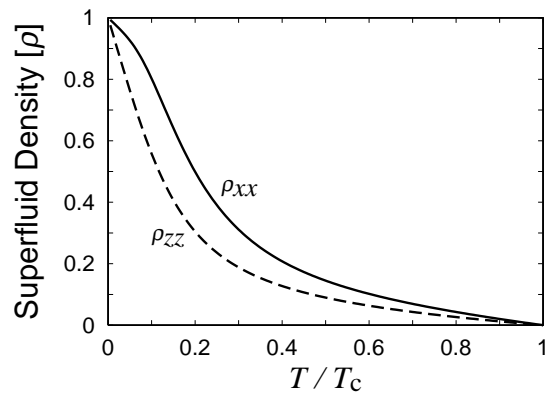}
\caption{
   The temperature dependence of
the superfluid densities
$\rho_{xx}$ and $\rho_{zz}$
for the Fermi velocity model, $v_{\rm I,II} \propto 1/N_{\rm I,II}$.
   $\nu=0.6$ and $\delta=0.9$.
}
\label{fig:6}
\end{figure}

   We have so far shown the results obtained
for the Fermi velocity model
of Eq.\ (\ref{eq:model-vF}) ($v_{\rm I} = v_{\rm II}$)
and Eq.\ (\ref{eq:C_I-II}) ($C_{\rm I,II}(\delta)=1 \pm \delta$).
   Alternatively, we have calculated
the same quantities
also for a different model: $v_{\rm I,II} \propto 1/N_{\rm I,II}$
$\bigl[$thus,
$v_{\rm II}=v_{\rm I}(1+\delta)/(1-\delta)$
instead of Eq.\ (\ref{eq:model-vF})$\bigr]$.
   The weighting factors are accordingly
$C_{\rm I}=1+\delta$
and $C_{\rm II}=(1+\delta)^2/(1-\delta)$ in this model,
and $\rho \equiv \rho_{xx}(T=0)=\rho_{zz}(T=0)
=(2Mv_{\rm I}^2N_0/3)(1+\delta)/(1-\delta)$.
   Because of these weighting factors
$C_{\rm I}=1+\delta$
and $C_{\rm II}=(1+\delta)^2/(1-\delta)$,
the larger $\delta$ ($-1<\delta<1$),
the more enhanced the contribution of the Fermi surface II becomes.
   It is an opposite dependence on $\delta$
as compared to the case of Eq.\ (\ref{eq:C_I-II})
($C_{\rm I,II}(\delta)=1 \pm \delta$).
   The results for the superfluid densities are qualitatively similar to
those obtained for the previous model of Eq.\ (\ref{eq:model-vF})
($v_{\rm I} = v_{\rm II}$)
with replacing $\delta \rightarrow -\delta$
in Figs.\ \ref{fig:2}--\ref{fig:5}.
   We plot, nevertheless, a result for an extreme case in Fig.\ \ref{fig:6},
where we set $\nu=0.6$ and $\delta=0.9$.
   In this case,
the superfluid densities are predominantly determined
by the contribution of the Fermi surface II with gap nodes,
owing to the extreme value $\delta=0.9$
and the resulting weighting factor $C_{\rm II}=(1+\delta)^2/(1-\delta)
\gg C_{\rm I}=1+\delta$.
   It is noticed in Fig.\ \ref{fig:6} that
the superfluid densities are suppressed at high temperatures.
   The temperature dependence of $\rho_{xx}$
in the region $T \le 0.3 T_{\mathrm c}$ of Fig.\ \ref{fig:6}
is well fitted into an experimental result for $1/\lambda_{\mathrm L}^2(T)$
in CePt$_3$Si,\cite{bonalde}
and an unusually strong suppression of $1/\lambda_{\mathrm L}^2(T)$
at high temperatures\cite{bonalde} is
somewhat similar to that of $\rho_{zz}$ in Fig.\ \ref{fig:6}.
   However, the difference in the density of states estimated
from an band calculation for CePt$_3$Si
is $|\delta|\sim 0.25$--$0.3$.\cite{samokhin}
   Therefore,
the strong suppression of $1/\lambda_{\mathrm L}^2(T)$ at high temperatures
observed in CePt$_3$Si (Ref.\ \onlinecite{bonalde})
remains to be accounted for at this moment.

\section{Summary}
   We calculated
the temperature dependence of
the superfluid densities
$\rho_{xx}(T)$ and $\rho_{zz}(T)$
for the noncentrosymmetric superconductor
with the Rashba-type spin-orbit coupling
represented by Eq.\ (\ref{eq:gk_SO}).
   We showed that
the gap function of Eq.\ (\ref{eq:OP-A2+s}),
which has the spin-singlet and spin-triplet pairing components,
explains the line-node-gap temperature dependence
of the experimentally observed $\lambda_{\mathrm L}(T)$
in CePt$_3$Si.\cite{bauer2,bonalde}
   While the low-temperature behavior ($T \alt 0.2 T_{\mathrm c}$) of
$1/\lambda_{\mathrm L}^2(T)$
can be reproduced qualitatively by that gap function,
the high-temperature one still
remains to be accounted for.

   The detailed information on the Fermi surfaces in the actual material
CePt$_3$Si is not available so far.\cite{hashimoto} 
The main difficulty here
lies in the fact that this material is
a heavy fermion system with strongly renormalized carriers. 
   If the Fermi velocity on the Fermi surface with gap nodes
(on the Fermi surface II in our assumption) is sufficiently large
in the case of the model $v_{\rm I,II} \propto 1/N_{\rm I,II}$
(otherwise, if $N_{\rm II} \gg N_{\rm I}$
in the case of $v_{\rm I} \simeq v_{\rm II}$),
the anomalously strong suppression of $1/\lambda_{\mathrm L}^2(T)$
observed experimentally at high temperatures\cite{bonalde}
may be explained as in Fig.\ \ref{fig:6}
(where $C_{\rm II} \gg C_{\rm I}$). Such an assumption seems not unreasonable in view
to the large renormalization factors for the effective mass suggested from
thermodynamic measurements.
On the other hand,
the London penetration depth
was not measured on a single crystal,
but on polycrystalline and powder samples,\cite{bauer2,bonalde}
to which the anomalous behavior at high temperatures
could be attributed.
   We also note that
the superconducting transition around $T_{\mathrm c}$
is rather broad in CePt$_3$Si at least at this moment.\cite{scheidt}
   Thus also an unusual behavior of the superconducting phase close
to $ T_c $ may play a role in the $T$-dependence of
 $1/\lambda_{\mathrm L}^2(T)$. 
   In any case, further experimental studies
(e.g., experimental measurements on a single crystal)
and theoretical studies using more information on the
Fermi surfaces involved in superconductivity are needed in the future
to accomplish a detailed fitting.
It would be also important to test experimentally
CePt$_3$Si for the
intriguing nonmonotonic temperature dependence
of $ \rho_{zz}/\rho_{xx} $ (see Fig.\ \ref{fig:5}),
which could provide information on effective parameters of the model.

\begin{acknowledgments}
   We would like to thank 
D.~F.~Agterberg, I.~Bonalde, E.~Bauer, J.~Goryo, Y.~Kato,
M.~Matsumoto, M.~Yogi, K.~Izawa, and A.~Koga
for helpful discussions.
   One of us (N.H.) acknowledges support from
the Japan Society for the Promotion of Science (2003 PFRA Program). 
   We are also grateful for financial support
from the Swiss Nationalfonds and the NCCR MaNEP.
\end{acknowledgments}

\appendix*
\section{}
   In this Appendix, we describe the procedure for deriving
the quasiclassical Green functions
for a noncentrosymmetric superconductor
with the Rashba-type spin-orbit coupling
represented by Eqs.\ (\ref{eq:SO}) and (\ref{eq:gk_SO})
and with the gap function of Eq.\ (\ref{eq:OP-A2+s}).
   The explicit form of the Eilenberger equations which will be given here
$\bigl[$Eqs.\ (\ref{eilen-A2-a}) and (\ref{eilen-A2-d})$\bigr]$
would be useful for future studies in inhomogeneous systems
such as surfaces, junctions, and vortices\cite{hayashi05-2,hayashi05-3}
in the noncentrosymmetric superconductor.

   We start with the Eilenberger equation 
given in Eq.\ (\ref{eq:eilen0}), namely\cite{eq-eilen}
\begin{equation}
i \partial {\check g}
+ \bigl[ i\omega_n {\check \tau}_{3}
-\alpha {\check {\bm g}}_k \cdot {\check {\bm S}}
-{\check \Delta},
{\check g} \bigr]
=0,
\label{eq:eilen0-A}
\end{equation}
where we have defined
$\partial = {\bm v}_{\mathrm F}(s) \cdot {\bm \nabla}$.
   It is convenient to define,
for the Green functions $\bigl[$Eq.\ (\ref{eq:qcg})$\bigr]$,
\begin{eqnarray}
{\hat g}={\hat g}_0,
\quad
{\hat f}={\hat f}_0 i{\hat \sigma}_y,
\quad
{\hat {\bar f} }=-i{\hat \sigma}_y {\hat {\bar f} }_0,
\quad
{\hat {\bar g} } =
  -{\hat \sigma}_y {\hat {\bar g} }_0 {\hat \sigma}_y.
\nonumber \\
\label{eq:Green-A}
\end{eqnarray}
   We also define,
for the gap function $\bigl[$Eq.\ (\ref{eq:OP-A2+s})$\bigr]$,
\begin{eqnarray}
{\hat \Delta}={\hat \Delta }_0 i{\hat \sigma}_y,
\quad
{\hat \Delta}^\dagger=-i{\hat \sigma}_y {\hat \Delta}_0^\dagger,
\end{eqnarray}
where
\begin{eqnarray}
{\hat \Delta }_0({\bm r}, {\tilde {\bm k}})
=
\Psi({\bm r}) {\hat \sigma}_0
 +
\Delta({\bm r})
\bigl(
  - {\tilde k}_y {\hat \sigma}_x
  + {\tilde k}_x {\hat \sigma}_y
\bigr).
\label{eq:Delta-A}
\end{eqnarray}
   The Eilenberger equation (\ref{eq:eilen0-A})
is then written down to
\begin{subequations}
\label{eq:eilen2}
\begin{eqnarray}
& &   \nonumber
\partial {\hat g}_0
+i\alpha {\bm g}_k \cdot
\bigl(
   {\hat {\bm \sigma}} {\hat g}_0
   - {\hat g}_0 {\hat {\bm \sigma}}
\bigr)
\\ & & \qquad \qquad \quad {}
+
\bigl(
   {\hat \Delta}_0 {\hat {\bar f}}_0
   - {\hat f}_0 {\hat \Delta}_0^\dagger
\bigr)
=0,
\label{eq:eilen2-a}
\end{eqnarray}
\begin{eqnarray}
& &   \nonumber
(\partial +2\omega_n) {\hat f}_0
+i\alpha {\bm g}_k \cdot
\bigl(
   {\hat {\bm \sigma}} {\hat f}_0
   - {\hat f}_0 {\hat {\bm \sigma}}
\bigr)
\\ & & \qquad \qquad \qquad \quad {}
+
\bigl(
   {\hat \Delta}_0 {\hat {\bar g}}_0
   - {\hat g}_0 {\hat \Delta}_0
\bigr)
=0,
\label{eq:eilen2-b}
\end{eqnarray}
\begin{eqnarray}
& &   \nonumber
(\partial -2\omega_n) {\hat {\bar f}}_0
+i\alpha {\bm g}_k \cdot
\bigl(
     {\hat {\bm \sigma}} {\hat {\bar f}}_0
   - {\hat {\bar f}}_0 {\hat {\bm \sigma}}
\bigr)
\\ & & \qquad \qquad \qquad \quad {}
+
\bigl(
   {\hat \Delta}_0^\dagger {\hat g}_0
   - {\hat {\bar g}}_0 {\hat \Delta}_0^\dagger
\bigr)
=0,
\label{eq:eilen2-c}
\end{eqnarray}
\begin{eqnarray}
& &   \nonumber
\partial {\hat {\bar g}}_0
+i\alpha {\bm g}_k \cdot
\bigl(
     {\hat {\bm \sigma}} {\hat {\bar g}}_0
   - {\hat {\bar g}}_0 {\hat {\bm \sigma}}
\bigr)
\\ & & \qquad \qquad \quad {}
+
\bigl(
     {\hat \Delta}_0^\dagger {\hat f}_0
   - {\hat {\bar f}}_0 {\hat \Delta}_0
\bigr)
=0,
\label{eq:eilen2-d}
\end{eqnarray}
\end{subequations}
where
${\bm g}_k
= \sqrt{3/2}(-{\tilde k}_y, {\tilde k}_x, 0 )$.

   From Eq.\ (\ref{eq:normalization}),
the normalization conditions are
\begin{subequations}
\label{normalization-A}
\begin{eqnarray}
{\hat g}_0^2
  + {\hat f}_0
    {\hat {\bar f}}_0
={\hat \sigma}_0,
\end{eqnarray}
\begin{eqnarray}
{\hat {\bar g}}_0^2
  + {\hat {\bar f}}_0
    {\hat f}_0
={\hat \sigma}_0,
\end{eqnarray}
\begin{eqnarray}
 {\hat g}_0
 {\hat f}_0
= -{\hat f}_0
   {\hat {\bar g}}_0,
\end{eqnarray}
\begin{eqnarray}
 {\hat {\bar g}}_0
 {\hat {\bar f}}_0
= -{\hat {\bar f}}_0
   {\hat g}_0.
\end{eqnarray}
\end{subequations}

   Now, we consider
 a rotation in the spin space which is represented by
the matrices
($
{\hat U}_k^\dagger {\hat U}_k
=
{\hat U}_k {\hat U}_k^\dagger
 = {\hat \sigma}_0
$),\cite{paolo3,fujimoto}
\begin{eqnarray}
{\hat U}_k
=
\frac{1}{\sqrt{2}}
\begin{pmatrix}
  1 & -{\bar k}'_+  \\
  {\bar k}'_-   & 1
\end{pmatrix},
\quad
{\hat U}_k^\dagger
=
\frac{1}{\sqrt{2}}
\begin{pmatrix}
  1 & {\bar k}'_+  \\
  -{\bar k}'_-   & 1
\end{pmatrix},
\label{eq:Operator-A2}
\end{eqnarray}
%
   where
${\bar {\mathbf k}}=({\bar k}_x, {\bar k}_y, 0)
=(\cos\phi, \sin\phi,0)$
and
${\bar k}'_\pm = {\bar k}_y \pm i{\bar k}_x$.
   A physical meaning of this rotation is that
after the rotation
the quantization axis in the spin space becomes parallel to
the vector ${\bm g}_k \parallel (-{\tilde k}_y, {\tilde k}_x, 0)$
at each position on the Fermi surface.\cite{bauer2,paolo3,saxena}
   We define the matrix elements
of the Green functions
after the rotation as
\begin{eqnarray}
{\hat U}^\dagger_k {\hat g}_0 {\hat U}_k
=
\begin{pmatrix}
 g_{a} & g_{b}  \\
 g_{c} & g_{d}
\end{pmatrix},
\quad
{\hat U}^\dagger_k {\hat f}_0 {\hat U}_k
=
\begin{pmatrix}
 f_{a} & f_{b}  \\
 f_{c} & f_{d}
\end{pmatrix},
\nonumber  \\
{\hat U}^\dagger_k {\hat {\bar f}}_0 {\hat U}_k
=
\begin{pmatrix}
 {\bar f}_{a} & {\bar f}_{b}  \\
 {\bar f}_{c} & {\bar f}_{d}
\end{pmatrix},
\quad
{\hat U}^\dagger_k {\hat {\bar g}}_0 {\hat U}_k
=
\begin{pmatrix}
 {\bar g}_{a} & {\bar g}_{b}  \\
 {\bar g}_{c} & {\bar g}_{d}
\end{pmatrix}.
\label{eq:G-abcd1}
\end{eqnarray}
   We apply
${\hat U}^\dagger_k$ and ${\hat U}_k$
to the Eilenberger equations (\ref{eq:eilen2})
from left and right respectively,
so that
we obtain the following sets of equations
in the case when
${\bm g}_k$ and ${\hat \Delta}_0$ are given
in Eqs.\ (\ref{eq:gk_SO}) and (\ref{eq:Delta-A}).
   Here, we define $\alpha' = \alpha\sqrt{3/2}$ and
\begin{eqnarray}
\Delta_{\rm I} = \Psi+\Delta\sin\theta,
\quad
\Delta_{\rm II} = \Psi-\Delta\sin\theta.
\end{eqnarray}
For the Green functions with the suffix {\it ``a"},
\begin{subequations}
\label{eilen-A2-a}
\begin{eqnarray}
\partial g_{a}
+
\Delta_{\rm II} {\bar f}_{a}
-
\Delta_{\rm II}^* f_{a}
=0,
\end{eqnarray}
\begin{eqnarray}
(\partial +2\omega_n) f_{a}
+
\Delta_{\rm II} {\bar g}_{a}
-
\Delta_{\rm II} g_{a}
=0,
\end{eqnarray}
\begin{eqnarray}
(\partial -2\omega_n) {\bar f}_{a}
+
\Delta_{\rm II}^* g_{a}
-
\Delta_{\rm II}^* {\bar g}_{a}
=0,
\end{eqnarray}
\begin{eqnarray}
\partial {\bar g}_{a}
+
\Delta_{\rm II}^* f_{a}
-
\Delta_{\rm II} {\bar f}_{a}
=0.
\end{eqnarray}
\end{subequations}
%
   For the {\it ``b"},
\begin{subequations}
\label{eilen-A2-b}
\begin{eqnarray}
\partial g_{b}
-2i
\alpha'\sin\theta
g_{b}
+
\Delta_{\rm II} {\bar f}_{b}
-
\Delta_{\rm I}^* f_{b}
=0,
\end{eqnarray}
\begin{eqnarray}
(\partial +2\omega_n) f_{b}
-2i
\alpha'\sin\theta
f_{b}
+
\Delta_{\rm II} {\bar g}_{b}
-
\Delta_{\rm I} g_{b}
=0,
\end{eqnarray}
\begin{eqnarray}
(\partial -2\omega_n) {\bar f}_{b}
-2i
\alpha'\sin\theta
{\bar f}_{b}
+
\Delta_{\rm II}^* g_{b}
-
\Delta_{\rm I}^* {\bar g}_{b}
=0,
\end{eqnarray}
\begin{eqnarray}
\partial {\bar g}_{b}
-2i
\alpha'\sin\theta
{\bar g}_{b}
+
\Delta_{\rm II}^* f_{b}
-
\Delta_{\rm I} {\bar f}_{b}
=0.
\end{eqnarray}
\end{subequations}
%
   For the {\it ``c"},
\begin{subequations}
\label{eilen-A2-c}
\begin{eqnarray}
\partial g_{c}
+2i
\alpha'\sin\theta
g_{c}
+
\Delta_{\rm I} {\bar f}_{c}
-
\Delta_{\rm II}^* f_{c}
=0,
\end{eqnarray}
\begin{eqnarray}
(\partial +2\omega_n) f_{c}
+2i
\alpha'\sin\theta
f_{c}
+
\Delta_{\rm I} {\bar g}_{c}
-
\Delta_{\rm II} g_{c}
=0,
\end{eqnarray}
\begin{eqnarray}
(\partial -2\omega_n) {\bar f}_{c}
+2i
\alpha'\sin\theta
{\bar f}_{c}
+
\Delta_{\rm I}^* g_{c}
-
\Delta_{\rm II}^* {\bar g}_{c}
=0,
\end{eqnarray}
\begin{eqnarray}
\partial {\bar g}_{c}
+2i
\alpha'\sin\theta
{\bar g}_{c}
+
\Delta_{\rm I}^* f_{c}
-
\Delta_{\rm II} {\bar f}_{c}
=0.
\end{eqnarray}
\end{subequations}
%
   For the {\it ``d"},
\begin{subequations}
\label{eilen-A2-d}
\begin{eqnarray}
\partial g_{d}
+
\Delta_{\rm I} {\bar f}_{d}
-
\Delta_{\rm I}^* f_{d}
=0,
\end{eqnarray}
\begin{eqnarray}
(\partial +2\omega_n) f_{d}
+
\Delta_{\rm I} {\bar g}_{d}
-
\Delta_{\rm I} g_{d}
=0,
\end{eqnarray}
\begin{eqnarray}
(\partial -2\omega_n) {\bar f}_{d}
+
\Delta_{\rm I}^* g_{d}
-
\Delta_{\rm I}^* {\bar g}_{d}
=0,
\end{eqnarray}
\begin{eqnarray}
\partial {\bar g}_{d}
+
\Delta_{\rm I}^* f_{d}
-
\Delta_{\rm I} {\bar f}_{d}
=0.
\end{eqnarray}
\end{subequations}
   Note above that
the Green functions with the suffixes
{\it ``a"}, {\it ``b"}, {\it ``c"}, and {\it ``d"}
are decoupled from each other
in these sets of the Eilenberger equations.

    We can solve Eqs.\ (\ref{eilen-A2-b}) and (\ref{eilen-A2-c})
in the case of spatially uniform system,
and find that
$g_{b}=f_{b}={\bar f}_{b}={\bar g}_{b}=0$
and
$g_{c}=f_{c}={\bar f}_{c}={\bar g}_{c}=0$
for $\alpha' \neq 0$.
   We also notice that
the Green functions
with the suffixes {\it ``b"} and {\it ``c"}
are zero everywhere
(even if the order parameters
$\Delta_{\rm I,II}$ are spatially inhomogeneous),
when solving the differential equations
(\ref{eilen-A2-b}) and (\ref{eilen-A2-c})
with the initial values equal to zero.
   On the other hand,
the Green functions with the suffixes
{\it ``a"} and {\it ``d"}
in Eqs.\ (\ref{eilen-A2-a}) and (\ref{eilen-A2-d})
have finite values in general.
   Therefore,
we can rewrite Eq.\ (\ref{eq:G-abcd1}) as
\begin{eqnarray}
{\hat U}^\dagger_k {\hat g}_0 {\hat U}_k
=
\begin{pmatrix}
 g_{a} & 0  \\
 0 & g_{d}
\end{pmatrix},
\quad
{\hat U}^\dagger_k {\hat f}_0 {\hat U}_k
=
\begin{pmatrix}
 f_{a} & 0  \\
 0 & f_{d}
\end{pmatrix},
\nonumber  \\
{\hat U}^\dagger_k {\hat {\bar f}}_0 {\hat U}_k
=
\begin{pmatrix}
 {\bar f}_{a} & 0  \\
 0 & {\bar f}_{d}
\end{pmatrix},
\quad
{\hat U}^\dagger_k {\hat {\bar g}}_0 {\hat U}_k
=
\begin{pmatrix}
 {\bar g}_{a} & 0  \\
 0 & {\bar g}_{d}
\end{pmatrix}.
\end{eqnarray}
   We obtain accordingly
\begin{subequations}
\begin{eqnarray}
{\hat g}_0
=
\frac{1}{2}
\begin{pmatrix}
 g_{d}+g_{a} &
 -{\bar k}'_+ (g_{d}-g_{a})  \\
 -{\bar k}'_- (g_{d}-g_{a})  &
 g_{d}+g_{a}
\end{pmatrix},
\end{eqnarray}

\begin{eqnarray}
{\hat f}_0
=
\frac{1}{2}
\begin{pmatrix}
 f_{d}+f_{a} &
 -{\bar k}'_+ (f_{d}-f_{a})  \\
 -{\bar k}'_- (f_{d}-f_{a})  &
 f_{d}+f_{a}
\end{pmatrix},
\end{eqnarray}

\begin{eqnarray}
{\hat {\bar f}}_0
=
\frac{1}{2}
\begin{pmatrix}
 {\bar f}_{d}+{\bar f}_{a} &
 -{\bar k}'_+ ({\bar f}_{d}-{\bar f}_{a})  \\
 -{\bar k}'_- ({\bar f}_{d}-{\bar f}_{a})  &
 {\bar f}_{d}+{\bar f}_{a}
\end{pmatrix},
\end{eqnarray}

\begin{eqnarray}
{\hat {\bar g}}_0
=
\frac{1}{2}
\begin{pmatrix}
 {\bar g}_{d}+{\bar g}_{a} &
 -{\bar k}'_+ ({\bar g}_{d}-{\bar g}_{a})  \\
 -{\bar k}'_- ({\bar g}_{d}-{\bar g}_{a})  &
 {\bar g}_{d}+{\bar g}_{a}
\end{pmatrix},
\end{eqnarray}
\end{subequations}
and finally, from Eq.\ (\ref{eq:Green-A}),
\begin{subequations}
\begin{eqnarray}
{\hat g}
&=&
g_{\rm I} {\hat \sigma}_{\rm I}
+
g_{\rm II} {\hat \sigma}_{\rm II},
  \\
{\hat f}
&=&
\bigl(
f_{\rm I} {\hat \sigma}_{\rm I}
+
f_{\rm II} {\hat \sigma}_{\rm II}
\bigr) i{\hat \sigma}_y,
  \\
{\hat {\bar f} }
&=&
-i{\hat \sigma}_y
\bigl(
 {\bar f}_{\rm I} {\hat \sigma}_{\rm I}
 +
 {\bar f}_{\rm II} {\hat \sigma}_{\rm II}
\bigr),  \\
{\hat {\bar g} }
&=&
-{\hat \sigma}_y
\bigl(
  {\bar g}_{\rm I} {\hat \sigma}_{\rm I}
  +
  {\bar g}_{\rm II} {\hat \sigma}_{\rm II}
\bigr)
{\hat \sigma}_y,
\end{eqnarray}
\end{subequations}
%
%
with 
\begin{eqnarray}
{\hat \sigma}_{\rm I,II} = \frac{1}{2}
\Bigl(
{\hat \sigma}_0
   \pm {\bar {\bm g}}_k \cdot {\hat {\bm \sigma}}
\Bigr),
\quad
{\bar {\bm g}}_k
= (-{\bar k}_y,{\bar k}_x,0).
\end{eqnarray}
   Here, we have defined
$A_{\rm I} \equiv A_{d}$ and $A_{\rm II} \equiv A_{a}$,
$\bigl($$A=\{g,f,{\bar f},{\bar g}\}$$\bigr)$.

   In the case of spatially uniform system,
we obtain
from the Eilenberger equations (\ref{eilen-A2-a}) and (\ref{eilen-A2-d})
and the normalization conditions in Eq.\ (\ref{normalization-A}),
\begin{subequations}
\begin{eqnarray}
g_{\rm I}
\equiv
g_{d}
=
    \frac{ \omega_n }{B_{\rm I}},
\quad
g_{\rm II}
\equiv
g_{a}
=
    \frac{ \omega_n }{B_{\rm II}},
\end{eqnarray}

\begin{eqnarray}
f_{\rm I}
\equiv
f_{d}
=
    \frac{ \Delta_{\rm I} }{B_{\rm I}},
\quad
f_{\rm II}
\equiv
f_{a}
=
    \frac{ \Delta_{\rm II} }{B_{\rm II}},
\label{eq:f_I-II}
\end{eqnarray}

\begin{eqnarray}
{\bar f}_{\rm I}
\equiv
{\bar f}_{d}
=
\frac{ \Delta_{\rm I}^* }{B_{\rm I}},
\quad
{\bar f}_{\rm II}
\equiv
{\bar f}_{a}
=
\frac{ \Delta_{\rm II}^* }{B_{\rm II}},
\end{eqnarray}

\begin{eqnarray}
{\bar g}_{\rm I}
\equiv
{\bar g}_{d}
=
\frac{ -\omega_n }{B_{\rm I}},
\quad
{\bar g}_{\rm II}
\equiv
{\bar g}_{a}
=
\frac{ -\omega_n }{B_{\rm II}},
\end{eqnarray}
\label{eq:A-uniform-G}
\end{subequations}
with
\begin{eqnarray}
B_{\rm I,II}
=
\sqrt{
      \omega_n^2 
      +|\Delta_{\rm I,II}|^2
    }.
\end{eqnarray}
At last,
in these equations
we replace $\omega_n \rightarrow \omega_n +q$
comparing Eq.\ (\ref{eq:eilen-SFD}) with Eq.\ (\ref{eq:eilen0-A}),
so that we obtain the Green functions in Eqs.\ (\ref{eq:g-elements}).




\begin{thebibliography}{00}

\bibitem{edelstein}
V. M. Edelstein,
Phys. Rev. Lett. {\bf 75}, 2004 (1995).

\bibitem{gorkov}
L. P. Gor'kov and E. I. Rashba,
Phys. Rev. Lett. {\bf 87}, 037004 (2001).

\bibitem{paolo1}
P. A. Frigeri, D. F. Agterberg, A. Koga, and M. Sigrist,
Phys. Rev. Lett. {\bf 92}, 097001 (2004);
{\bf 93}, 099903(E) (2004).

\bibitem{samokhin}
K. V. Samokhin, E. S. Zijlstra, and S. K. Bose,
Phys. Rev. B {\bf 69}, 094514 (2004);
{\bf 70}, 069902(E) (2004).

\bibitem{sergienko}
I. A. Sergienko and S. H. Curnoe,
Phys. Rev. B {\bf 70}, 214510 (2004).

\bibitem{sergienko2}
I. A. Sergienko,
Physica B {\bf 359-361}, 581 (2005).

\bibitem{bauer}
E. Bauer,
G. Hilscher, H. Michor, Ch. Paul, E. W. Scheidt, A. Gribanov,
Yu. Seropegin, H. No\"el, M. Sigrist, and P.~Rogl,
Phys. Rev. Lett. {\bf 92}, 027003 (2004).

\bibitem{bauer3}
E. Bauer,
G. Hilscher, H. Michor, M. Sieberer, E.~W.~Scheidt, A. Gribanov,
Yu. Seropegin, P. Rogl, A.~Amato, W. Y. Song, J.-G. Park,
D. T. Adroja, M.~Nicklas, G. Sparn, M. Yogi, and Y. Kitaoka,
Physica B {\bf 359-361}, 360 (2005).

\bibitem{bauer4}
E. Bauer,
G. Hilscher, H. Michor, M. Sieberer, E.~W.~Scheidt, A. Gribanov,
Yu. Seropegin, P. Rogl, W.~Y.~Song, J.-G. Park,
D. T. Adroja, A. Amato, M. Nicklas, G. Sparn, M. Yogi, and Y. Kitaoka,
Czech. J. Phys. Suppl. {\bf 54}, D401 (2004).

\bibitem{bauer2}
E. Bauer, I. Bonalde, and M. Sigrist,
Fiz. Nizk. Temp. {\bf 31}, 984 (2005),
[Low Temp. Phys. {\bf 31}, 748 (2005)].

\bibitem{dresselhaus}
G. Dresselhaus,
Phys. Rev. {\bf 100}, 580 (1955).

\bibitem{rashba}
E. I. Rashba,
Fiz. Tverd. Tela (Leningrad) {\bf 1}, 407 (1959),
[Sov. Phys. Solid State {\bf 1}, 366 (1959)]. 

\bibitem{yasuda}
T. Yasuda,
H. Shishido, T. Ueda, S. Hashimoto, R. Settai, T. Takeuchi, T. D. Matsuda,
Y. Haga, and Y. Onuki,
J.~Phys. Soc. Jpn. {\bf 73}, 1657 (2004).

\bibitem{yogi}
M. Yogi,
Y. Kitaoka, S. Hashimoto, T. Yasuda, R. Settai, T. D. Matsuda,
Y. Haga, Y. Onuki, P. Rogl, and E. Bauer,
Phys. Rev. Lett. {\bf 93}, 027003 (2004).

\bibitem{yogi2}
M. Yogi,
Y. Kitaoka, S. Hashimoto, T. Yasuda, R. Settai, T. D. Matsuda,
Y. Haga, Y. Onuki, P. Rogl, and E. Bauer,
Physica B {\bf 359-361}, 389 (2005).

\bibitem{yogi3}
M. Yogi,
Y. Kitaoka, S. Hashimoto, T. Yasuda, R. Settai, T. D. Matsuda,
Y. Haga, Y. Onuki, P. Rogl, and E. Bauer,
J. Phys. Chem. Solids (to be published).

\bibitem{ueda}
K. Ueda, K. Hamamoto, T. Kohara, G. Motoyama, and Y.~Oda,
Physica B {\bf 359-361}, 374 (2005).

\bibitem{izawa}
K. Izawa, Y. Kasahara, Y. Matsuda, K. Behnia, T.~Yasuda, R. Settai,
and Y. Onuki,
Phys. Rev. Lett. {\bf 94}, 197002 (2005);
K. Maki, D. Parker, and H. Won,
cond-mat/0508429.

\bibitem{bonalde}
I. Bonalde, W. Br\"amer-Escamilla, and E. Bauer,
Phys. Rev. Lett. {\bf 94}, 207002 (2005).

\bibitem{metoki}
N. Metoki,
K. Kaneko, T. D. Matsuda, A. Galatanu, T.~Takeuchi, S. Hashimoto,
T. Ueda, R. Settai, Y. Onuki, and N. Bernhoeft,
J. Phys.: Condens. Matter {\bf 16}, L207 (2004);
Physica B {\bf 359-361}, 383 (2005).

\bibitem{amato}
A. Amato, E. Bauer, and C. Baines,
Phys. Rev. B {\bf 71}, 092501 (2005).

\bibitem{ishikawa}
M. Ishikawa, S. Yamashita, Y. Nakazawa, N. Wada, and N. Takeda,
J. Phys.: Condens. Matter {\bf 17}, L231 (2005);
M.~Ishikawa and N. Takeda,
Solid State Commun. {\bf 133}, 249 (2005).

\bibitem{hayashi05-1}
N. Hayashi, K. Wakabayashi, P. A. Frigeri, and M. Sigrist,
cond-mat/0504176.

\bibitem{hayashi05-2}
N. Hayashi, K. Wakabayashi, P. A. Frigeri, Y. Kato, and M. Sigrist,
cond-mat/0510547.

\bibitem{samokhin2}
From an another approach,
the NMR relaxation rate in noncentrosymmetric superconductors
is theoretically discussed in,
K.~V.~Samokhin, Phys. Rev. B {\bf 72}, 054514 (2005);
see also S.~Fujimoto, Ref.\ \onlinecite{fujimoto}.

\bibitem{paolo2}
P. A. Frigeri, D. F. Agterberg, and M. Sigrist,
New J. Phys. {\bf 6}, 115 (2004).

\bibitem{paolo2-2}
P. A. Frigeri, D. F. Agterberg, A. Koga, and M. Sigrist,
Physica B {\bf 359-361}, 371 (2005).

\bibitem{kaur}
R. P. Kaur, D. F. Agterberg, and M. Sigrist,
Phys. Rev. Lett. {\bf 94}, 137002 (2005). 

\bibitem{paolo3}
P. A. Frigeri, D. F. Agterberg, I. Milat, and M. Sigrist,
cond-mat/0505108.

\bibitem{anderson}
P. W. Anderson,
Phys. Rev. B {\bf 30}, 4000 (1984).

\bibitem{d-vector}
One may choose
more complicated spin-triplet state with ${\bm d}_k \parallel {\bm g}_k$,
for example
${\bm d}_k
=\Delta (1+\cos 2\theta)
(-{\tilde k}_y,{\tilde k}_x,0)$.
However, in this paper we consider the simplest one,
${\bm d}_k
=\Delta
(-{\tilde k}_y,{\tilde k}_x,0)$,
for clarity.

\bibitem{fujimoto}
S. Fujimoto,
Phys. Rev. B {\bf 72}, 024515 (2005).

\bibitem{sergienko3}
I. A. Sergienko,
Phys. Rev. B {\bf 69}, 174502 (2004).

\bibitem{eilen}
G. Eilenberger,
Z. Phys. {\bf 214}, 195 (1968).

\bibitem{LO}
A. I. Larkin and Yu. N. Ovchinnikov,
Zh. \'Eksp. Teor. Fiz. {\bf 55}, 2262 (1968),
[Sov. Phys. JETP {\bf 28}, 1200 (1969)].

\bibitem{serene}
J. W. Serene and D. Rainer,
Phys. Rep. {\bf 101}, 221 (1983).

\bibitem{schopohl80}
N. Schopohl,
J. Low Temp. Phys. {\bf 41}, 409 (1980).

\bibitem{rieck}
C. T. Rieck, K. Scharnberg, and N. Schopohl,
J. Low Temp. Phys. {\bf 84}, 381 (1991).

\bibitem{choi}
C. H. Choi and J. A. Sauls,
Phys. Rev. B {\bf 48}, 13684 (1993).

\bibitem{kusunose}
H. Kusunose,
Phys. Rev. B {\bf 70}, 054509 (2004).

\bibitem{eq-eilen}
   In the beginning, in the Eilenberger equation in the general form of
Eqs.\ (\ref{eq:eilen0}) and (\ref{eq:eilen0-A}),
we assume that the spin-orbit coupling $\alpha$ is sufficiently weak
such that
the spin-degeneracy lifting is small and
the Fermi velocity $v_{\mathrm F}$ is constant with respect to
the spin projections.
   Then, after reaching finally
the physically relevant parts of the Eilenberger equation
$\bigl[$Eqs.\ (\ref{eilen-A2-a}) and (\ref{eilen-A2-d})$\bigr]$
for the specific set of the spin-orbit coupling and the gap function
$\bigl[$Eqs.\ (\ref{eq:gk_SO}) and (\ref{eq:OP-A2+s})$\bigr]$,
we can set $\alpha > T_{\mathrm c}$
(even $\alpha \gg T_{\mathrm c}$, but $\alpha \ll \varepsilon_{\mathrm F}$)
and relax the restriction on $v_{\mathrm F}$
because $v_{\mathrm F}$ in these
Eqs.\ (\ref{eilen-A2-a}) and (\ref{eilen-A2-d})
are interpreted as two mutually independent Fermi velocities
on the distinctly different Fermi surfaces split into I and II.

\bibitem{choi87}
C. H. Choi and P. Muzikar,
Phys. Rev. B {\bf 36}, 54 (1987).

\bibitem{choi88}
C. H. Choi and P. Muzikar,
Phys. Rev. B {\bf 37}, R5947 (1988).

\bibitem{choi89}
C. H. Choi and P. Muzikar,
Phys. Rev. B {\bf 39}, 11296 (1989).

\bibitem{muzikar95}
G. Preosti and P. Muzikar,
Phys. Rev. B {\bf 51}, R15634 (1995).

\bibitem{choi90}
C. H. Choi and P. Muzikar,
in {\it Field Theories in Condensed Matter Physics: A Workshop},
edited by Z. Tesanovic
(Addison-Wesley, California, 1990), p. 51.

\bibitem{mineev}
V. P. Mineev,
Int. J. Mod. Phys. B {\bf 18}, 2963 (2004).

\bibitem{gk}
Comparing the notation of the spin-triplet order parameter ($\Delta$)
with that in Ref.\ \onlinecite{paolo3} ($d$),
we have the relation $\sin\theta = d |{\bm g}_k| /\Delta
= |{\bm g}_k| \sqrt{2/3}$,
($\Delta = d\sqrt{3/2}$).
Substituting this into Eqs.\ (\ref{eq:rho}),
(\ref{eq:gap-singlet}), and (\ref{eq:gap-triplet}),
we can use these equations to calculate the superfluid density
also for other noncentrosymmetric systems with general ${\bm g}_k$
(but $\langle {\bm g}_k^2 \rangle$=1).

\bibitem{FS-SOC}
In principle, the anisotropy (or deformation) of each split Fermi surface
$\bigl($due to, {\it e.g.},
the anti-symmetric spin-orbit coupling
[Eq.\ (\ref{eq:SO})]$\bigr)$
can be incorporated
through ${\bm v}_{\mathrm F}(s)$
and $N_{\mathrm F}(s)$, namely
through the ${\tilde {\bm k}}$ dependence of
${\bm v}_{\rm I,II}$ in Eq.\ (\ref{eq:rho}) and
the forms of each Fermi-surface average $\langle \cdots \rangle$.
However, the most important qualitative effect of
those split Fermi surfaces characteristic of
the noncentrosymmetric systems is that different superconducting gaps
can appear on each Fermi surface.
Thermodynamic properties at low $T$ are determined
predominantly by the gap topology
(such as line-node, point-node, and full gaps).

\bibitem{yip}
S. K. Yip,
Phys. Rev. B {\bf 65}, 144508 (2002).

\bibitem{hashimoto}
S. Hashimoto,
T. Yasuda, T. Kubo, H. Shishido, T. Ueda, R. Settai,
T. D. Matsuda, Y. Haga, H. Harima, and Y.~Onuki,
J. Phys.: Condens. Matter {\bf 16}, L287 (2004);
Y.~Onuki,
T. Yasuda, H. Shishido, S. Hashimoto, T. Ueda, R. Settai,
T. D. Matsuda, Y. Haga, and H. Harima,
Physica B {\bf 359-361}, 368 (2005).

\bibitem{scheidt}
E.-W. Scheidt, F. Mayr, G. Eickerling, P. Rogl, and E.~Bauer,
J. Phys.: Condens. Matter {\bf 17}, L121 (2005);
See also,
J.~S.~Kim,
D. J. Mixson, D. J. Burnette, T. Jones, P. Kumar, B. Andraka, G. R. Stewart,
V. Craciun, W.~Acree,
H. Q. Yuan, D. Vandervelde, and M. B. Salamon,
Phys. Rev. B {\bf 71}, 212505 (2005).

\bibitem{hayashi05-3}
N. Hayashi, Y. Kato, K. Wakabayashi, P. A. Frigeri, and M. Sigrist,
cond-mat/0510548.

\bibitem{saxena}
S. S. Saxena and P. Monthoux,
Nature {\bf 427}, 799 (2004).

\end{thebibliography}
\end{document}